
\input amstex
\documentstyle{amsppt}
\rightheadtext{Invariant Integrability Criterion \dots}
\CenteredTagsOnSplits
\topmatter
\title
Invariant Integrability Criterion
for the Equations of Hydrodynamical Type.
\endtitle
\author
Pavlov M.V., Sharipov R.A., Svinolupov S.I.
\endauthor
\address
Landau Institute for Theoretical Physics, Kosigina street 2,
117940 Moscow, Russia;
Department of Mathematics, Bashkir State University, Frunze street 32,
450074 Ufa, Russia;
Institute of Mathematics, Chernishevsky street 112, 450000 Ufa,
Russia
\endaddress
\email
pavlov\@cpd.landau.free.net;
root\@bgua.bashkiria.su;
sersv\@nkc.bashkiria.su
\endemail
\thanks
This research was supported by Grant \#RK4000 from International
Science Foundation and by Grant \#93-011-16088 from Russian Fund for
Fundamental Researches.
\endthanks
\abstract
     Invariant integrability criterion for the equations of
hydrodynamical type is found. This criterion is written in the
form of vanishing for some tensor which is derived from the
velocities matrix of hydrodynamical equations.
\endabstract
\endtopmatter
\document
\head
1. Introduction.
\endhead
     Systems of quasilinear partial differential equations of
the first order arise in different models describing the motion
of continuous media. Special subclass of such systems is known
as a systems of equations of {\it hydrodynamical type}. In
spatially one-dimensional case they are written as follows
$$
u^i_t=\sum^n_{j=1}A^i_j(\bold u) u^j_x
\text{, \ where } i=1,\dots,n\tag1.1
$$
Among the systems \thetag{1.1} man can consider special
subclass of systems possessing the {\it Riemann invariants}.
These are the systems which can be transformed to the diagonal
form
$$
u^i_t=\lambda_i(\bold u) u^i_x
\text{, \ where } i=1,\dots,n\tag1.2
$$
by means of so called {\it point transformations}
$$
\tilde u^i=\tilde u^i(u^1,\dots,u^n)
\text{, \ where } i=1,\dots,n\tag1.3
$$
System of equations \thetag{1.1} is called the {\it
hydrodynamically integrable system} if it has the continuous
set of hydrodynamical symmetries (or hydrodynamical conservation
laws) parameterized by $n$ arbitrary functions of one variable.
For diagonal systems \thetag{1.2} with mutually distinct
characteristic velocities ($\lambda_i\neq\lambda_j$) one has the
well-developed theory of integration (see reviews \cite{1} and
\cite{2}). As it was shown in \cite{2} the diagonal system
\thetag{1.2} is integrable if and only if the following condition
is satisfied
$$
\partial_i\left(\frac{\partial_j\lambda_k}{\lambda_j-\lambda_k}
\right)= \partial_j\left(\frac{\partial_i\lambda_k}{\lambda_i-
\lambda_k}\right) \text{, \ where } i\neq k\neq j\tag1.4
$$
where $\partial_i=\partial/\partial u^i$ and $\partial_j=
\partial/\partial u^j$. Such systems are called {\it
semi-Hamiltonian systems}, for the property \thetag{1.4} itself
in Russian papers the term {\it semihamiltonity} is used. When
the diagonal system \thetag{1.2} possess this property it can
be integrated by means of "generalized hodograph method"
(details see in \cite{2}) \par
     In section 2 of this paper we consider the problem of
hydrodynamical integrability for the systems of equations
\thetag{1.1} with the velocities matrix $A^i_j(\bold u)$ of
general position i.e. eigenvalues of which are mutually distinct.
There we managed to prove the following fact: each
hydrodynamically integrable system \thetag{1.1} with the matrix
of general position is necessarily diagonalizable \footnote{May
be this is not new fact but we couldn't find it anywhere.}.
Summing up this fact with the results of \cite{2} we may state
the following theorem \par
\proclaim{Theorem 1} System of equations \thetag{1.1} with
the matrix of general position is hydrodynamically integrable
if and only if it is diagonalizable and semi-Hamiltonian.
\endproclaim
     Theorem 1 shows that the study of diagonal equations
\thetag{1.2} is very important. But it doesn't exclude the
necessity of study of general equations \thetag{1.1}. Indeed
the integrability test for the equations \thetag{1.1} of
general position according to the theorem 1 should include
3 steps
\roster
\item test of diagonalizability,
\item diagonalization by means of the point transformation
      \thetag{1.3},
\item test of semihamiltonity \thetag{1.4}.
\endroster
First of these steps is implemented by means of invariant
geometrical criterion from \cite{3}. This criterion consists in
vanishing of Haantjes's tensor derived from the velocities matrix
of the equations in question. This test was first applied to
the equations \thetag{1.1} in \cite{4}. \par
     Next step is an actual diagonalization. To pass this step
one should calculate the eigenvalues of the matrix
$A^i_j(\bold u)$, find its eigenvectors, properly normalize them
and then one should solve some system of ordinary differential
equations defining the transformation \thetag{1.2}. Because of
this step the total integrability test may be absolutely
inefficient since the case when the system of differential
equations is explicitly solvable is very rare event. However
if we find such solution the third step may have only the
difficulties in calculations. \par
     The presence of nonefficient step in the above integrability
test of the equations \thetag{1.1} is due to the absence of the
of invariant criterion for testing the semihamiltonity for
these equations. The main goal of this paper is to eliminate this
essential fault of the theory of such equations. As it was noted
in \cite{1}: It was Riemann who first recognized that the theory
of the equations \thetag{1.1} is the theory of tensors since the
components of matrix $A^i_j(\bold u)$ are transformed as the
components of tensor under the point transformations
\thetag{1.3}. Therefore it is natural to expect that the
semihamiltonity relationships \thetag{1.4} can be rewritten in
an invariant tensorial form. In the section 4 of this paper
we construct the tensor, vanishing of which is equivalent to
\thetag{1.4}. So invariant integrability criterion for the
equations of hydrodynamical type is obtained. New integrability
test now is absolutely efficient. It includes two steps
\roster
\item test of vanishing the tensor of Haantjes,
\item test of vanishing the semihamiltonity tensor.
\endroster \par
In order to construct the semihamiltonity tensor we use the
theory developed by Froelicher and Nijenhuis in \cite{5} and
\cite{6}. This theory in brief is given in section 3. According
to the theory of Froelicher and Nijenhuis each smooth manifold
is equipped with some Lie superalgebra of tensor fields of type
$(1,p)$. Note that the paper \cite{6} was written in 1956 but
unfortunately it wasn't known to specialists since for example
in \cite{7} the paper of Martin of 1959 is quoted as the first
paper in supermathematics. \par
     Authors are grateful to V.E.Adler and I.Yu.Cherdantzev for
the fruitful discussions. They are also grateful to
B.I.Suleymanov for supporting interest to this paper.
\head
2. Hydrodynamical integrability.
\endhead
     Let's take the system of equations of hydrodynamical type
\thetag{1.1} and let's add to it another such system with the
dynamics by the variable $\tau$ \par
$$
u^i_\tau=\sum^n_{j=1}B^i_j(\bold u) u^j_x \text{, \ where }
i=1,\dots,n\tag2.1
$$
\definition{Definition 1} System of equations \thetag{2.1}
is called the hydrodynamical symmetry for the equations
\thetag{1.1} if the equations \thetag{1.1} and \thetag{2.1} are
compatible. \enddefinition
     Now we shall study the question about the existence and the
number of hydrodynamical symmetries for the system of equations
\thetag{1.1}. Let the equations \thetag{1.1} and \thetag{2.1}
be compatible. Their compatibility conditions are written in
form of the relationships
$$
\align
&\sum^n_{s=1} A^i_s B^s_j=\sum^n_{s=1} B^i_s A^s_j \tag2.2\\
\split
&\sum^n_{s=1}
\bigl(\partial_s A^k_i B^s_j +
\partial_s A^k_j B^s_i +
\partial_i B^s_j A^k_s +
\partial_j B^s_i A^k_s\bigr)=
\hskip -2em\\
&=\sum^n_{s=1}\bigl(
\partial_s B^k_i A^s_j +
\partial_s B^k_j A^s_i +
\partial_i A^s_j B^k_s +
\partial_j A^s_i B^k_s\bigr)\hskip -2em
\endsplit \tag2.3
\endalign
$$
The relationship \thetag{2.2} means that the matrices the systems
\thetag{1.1} and \thetag{2.1} are commuting
$$
\bold A\bold B=\bold B\bold A \tag2.4
$$
Second relationship \thetag{2.3} also can be written in an invariant
form. In order to do it let's contract \thetag{2.3} with $X^i X^j$,
where $X^1,\dots,X^n$ are the components of some arbitrary vector
field $\bold X$.
$$
[\bold A\bold X,\bold B\bold X] - \bold A[\bold X,\bold B\bold X]
- \bold B[\bold A\bold X,\bold X] = 0\tag2.5
$$
This  result can be stated as a theorem. \par
\proclaim{Theorem 2} System of the equations \thetag{2.1} is a
hydrodynamical symmetry for the system \thetag{1.1} if and only
if for any vector field $\bold X$ the relationships \thetag{2.4}
and \thetag{2.5} hold. \endproclaim
     Let the operator field $\bold A = \bold A(\bold u)$ from
\thetag{1.1} have $n$ mutually distinct eigenvalues $\lambda_i =
\lambda_i(\bold u)$. Through $\bold X_1 ,\dots,\bold X_n$ we
denote the frame formed by eigenvectors of operator $\bold A$.
The choice of eigenvectors is not unique, there is the gauge
arbitrariness in $n$ scalar factors
$$
\bold X_i(\bold u) \longrightarrow f_i(\bold u) \bold X_i
(\bold u)\text{, \ where } f_i\neq 0 \tag2.6
$$
For the sake of brevity we introduce the following notations for
the Lie derivatives along the vector fields $\bold X_1,\dots,
\bold X_n$
$$
L_i=L_{\bold X_{\ssize i}} \tag{2.7}
$$
Mutual commutators of the vector fields $\bold X_1,\dots,
\bold X_n$ are convenient to be expanded in the frame formed
by these fields
$$
L_i\bold X_j=[\bold X_i,\bold X_j] =
\sum^n_{k=1}c^k_{ij}\bold X_k \tag{2.8}
$$
Parameters $c^k_{ij}=c^k_{ij}(\bold u)$ in \thetag{2.8} are to
be called the {\it structural scalars} of the frame $\bold X_1,
\dots, \bold X_n$. The term {\it structural constants} doesn't
suit since $c^k_{ij}$ depend on the point $\bold u$. \par
     From the algebra we know that the matrix $\bold B$ is
commuting with the matrix $\bold A$ having mutually distinct
eigenvalues then these two matrices are simultaneously
diagonalized in the frame $\bold X_1,\dots,\bold X_n$. Therefore
any operator $\bold B$ satisfying \thetag{2.4} is completely
defined by its eigenvalues $\mu_i=\mu_i(\bold u)$. Note that
from \thetag{2.5} we have the relationship
$$
\align
&[\bold A\bold X,\bold B\bold Y] +
[\bold A\bold Y,\bold B\bold X]
- \bold A[\bold X,\bold B\bold Y] - \\
&- \bold A[\bold Y,\bold B\bold X]
- \bold B[\bold A\bold X,\bold Y]
- \bold B[\bold A\bold Y,\bold X] = 0
\endalign
$$
which holds for two arbitrary vector fields $\bold X$ and
$\bold Y$. Let's substitute $\bold X=\bold X_i$ and $\bold Y=
\bold X_j$ into the above relationship. As a result we obtain
that it is equivalent to the pair of sets of relationships.
First set is algebraic with respect to the eigenvalues $\mu_i$
of the matrix $\bold B$
$$
\aligned
&c^i_{jk}(\lambda_j-\lambda_k)\mu_i +
c^i_{jk}(\lambda_k-\lambda_i)\mu_j + \\
&\quad + c^i_{jk}(\lambda_i-\lambda_j)\mu_k = 0
\text{, \ for } i\neq j,j\neq k,k\neq i
\endaligned \tag2.9
$$
Second set contains the partial differential equations with
respect to $\mu_i$
$$
L_i\mu_j=\lambda_{ij}\frac{\mu_i-\mu_j}
{\lambda_i-\lambda_j}\text{, \ for } i\neq j \tag2.10
$$
Here and everywhere below we use the notations $\lambda_{ij} =
L_i\lambda_j$ in terms of Lie derivatives \thetag{2.7}. \par
     System of differential equations \thetag{2.10} is
overdetermined. When it is compatible the maximal degree
of arbitrariness for its solutions is $n$ functions of one
variables. Let this degree of arbitrariness be actually realized.
Then
$$
\mu_i = \mu_i(f_1,\dots,f_n,\bold u)\tag2.11
$$
Let's substitute \thetag{2.11} into \thetag{2.9}. This leads to
the functional dependence for the parameters $f_1(z_1),\dots,
f_n(z_n)$ from \thetag{2.11}, which contradicts their
arbitrariness. Therefore the relationships \thetag{2.9} should
be trivial. For the structural scalars \thetag{2.8} this gives
$$
c^k_{ij} = 0 \text{, \ for } i\neq j, j\neq k, k\neq i
\tag2.12
$$
The relationships \thetag{2.12} have the important consequences
which are due to the following lemma. \par
\proclaim{Lemma 1} The linear operator of the general position
$\bold A = \bold A(\bold u)$ is diagonalizable by means of the
transformation \thetag{1.3} if and only if the relationships
\thetag{2.12} hold for the frame of its eigenvectors.
 \endproclaim
     We give the sketch of proof of this lemma. Operator
$\bold A(\bold u)$ is diagonal in the frame of its eigenvectors.
For $\bold A(\bold u)$ to be diagonalizable by the
transformation \thetag{1.3} this frame should be the coordinate
frame i.e. structural scalars of this frame should be identically
zero $c^k_{ij}=0$. The relationship \thetag{2.12} provides
vanishing most of these scalars. One can reach vanishing the
rest of these scalars by use of the gauge arbitrariness
\thetag{2.6}. \par
     Because of lemma 1 the further analysis of the compatibility
conditions for the equations \thetag{2.10} becomes unnecessary.
For the diagonal systems \thetag{1.2} such analysis was done by
S.P.Tsarev in \cite{2}. Note only that as result of such analysis
we add to \thetag{2.12} the following relationships
$$
\aligned
& L_i\left(\frac{\lambda_{jk}}{\lambda_j-\lambda_k}\right) -
  L_j\left(\frac{\lambda_{ik}}{\lambda_i-\lambda_k}\right) +\\
&+\frac{c^j_{ji}\lambda_{jk}}{\lambda_j-\lambda_k} -
  \frac{c^i_{ij}\lambda_{ik}}{\lambda_i-\lambda_k} = 0
\endaligned\tag2.13
$$
which hold for $i\neq j, j\neq k, k\neq i$. The relationships
\thetag{2.13} are the same as the semihamiltonity relationships
\thetag{1.4} but written in the frame of eigenvectors of
diagonalizable operator $\bold A(\bold u)$. The above
considerations prove the theorem 1 in the following form. \par
\proclaim{Theorem 3} System of the equations \thetag{1.1} with
the matrix of general position possess the continuous set of
hydrodynamical symmetries with functional arbitrariness given
by $n$ functions of one variable if and only if it is
diagonalizable and semi-Hamiltonian. \endproclaim
     Now let's study the similar question about the conservation
laws for \thetag{1.1}. On the set of their solutions we define
the integral functionals of the following form
$$
F=\int f(\bold u) dx\tag2.14
$$
Functional \thetag{2.14} is called the hydrodynamical
conservation law or the first integral for the equations
\thetag{1.1} if $\dot F=0$ when time derivative $\dot F$ is
calculated according to the dynamics given by \thetag{1.1}.
This derivative is the following integral functional
$$
G=\dot F=\int\left(\sum^n_{i=1}\sum^n_{j=1}
\partial_if A^i_j u^j_x\right) dx\tag2.15
$$
{}From vanishing the functional \thetag{2.15} we get the vanishing
of its variational derivatives
$$
\frac{\delta G}{\delta u^i} =
\sum^n_{j=1}\sum^n_{k=1}\bigl(\partial_i(\partial_kf A^k_j) -
\partial_j(\partial_k A^k_i)\bigr) u^j_x = 0
$$
This leads to the following relationship for the density
$f(\bold u)$ of the primary functional \thetag{2.14}
$$
\sum^n_{k=1}\bigl(\partial_i(\partial_kf A^k_j) -
\partial_j(\partial_k A^k_i)\bigr) = 0\tag2.16
$$
It is equivalent to the existence of the function $T(\bold u)$
such that
$$
G=\int\sum^n_{i=1}\sum^n_{j=1}\bigl(\partial_if A^i_j u^j_x
\bigr) dx = \int\frac{\partial T}{\partial x} dx \tag2.17
$$
Because of \thetag{2.17} the condition \thetag{2.16} is exactly
the condition of vanishing the functional $G=\dot F=0$. In order
to write the relationships \thetag{2.16} in an invariant form
let's choose two arbitrary vector fields $\bold X$ and
$\bold Y$. After contracting \thetag{2.16} with $X^i$ and $Y^j$
we can write the result of such contraction through the Lie
derivatives
$$
\bigl(L_{\bold X} L_{\bold A\bold Y} - L_{\bold Y} L_{\bold A
\bold X} - L_{\bold A[\bold X,\bold Y]}\bigr) f = 0\tag2.18
$$
This result can be stated in form of the following theorem.
\par
\proclaim{Theorem 4} Integral functional \thetag{2.14} is
hydrodynamical conservation law for the system of equations
\thetag{1.1} if and only if for any choice of vector fields
$\bold X$ and $\bold Y$ the equations \thetag{2.18} hold.
\endproclaim
     The relationship \thetag{2.18} is the system of differential
equations with respect to the unknown function $f(\bold u)$. One
should investigate it for  the  compatibility.  Let's  denote
$L_i  f=\varphi_i$.   Substituting   frame   vectors   $\bold
X_1,\dots,
\bold X_n$ for $\bold X$ and $\bold Y$ into the relationships
\thetag{2.18} we get the following equations
$$
L_i\varphi_j=\sum^n_{k=1} B^k_{ij}\varphi_k
\text{, \ for } i\neq j\tag2.19
$$
where the functions $B^k_{ij}$ are defined by the formulae
$$
B^k_{ij} = c^k_{ij}\frac{\lambda_k-\lambda_i}
{\lambda_j-\lambda_i} + \frac{\lambda_{ji}\delta^k_i}
{\lambda_j-\lambda_i} - \frac{\lambda_{ij}\delta^k_j}
{\lambda_j-\lambda_i}\tag2.20
$$
The equations \thetag{2.19} are analogous to the equations
\thetag{2.10}. When they are compatible their solutions have
the arbitrariness in $n$ functions of one variable. Let such
arbitrariness be actually realized. We look for the differential
consequences of the equations \thetag{2.19}. Among them we
find the following relationships
$$
B^i_{jk} L_i\varphi_i - B^j_{ik} L_j\varphi_j -
c^k_{ij} L_k\varphi_k=- \sum^n_{q=1} R^q_{kij}\varphi_q
\tag2.21
$$
which hold for $i\neq j, j\neq k, k\neq i$.
The values of $R^q_{kij}$ for $i\neq j, j\neq k, k\neq i$ are
calculated according to the formula
$$
\aligned
R^q_{kij}&=L_iB^q_{jk}+\sum^n_{s\neq i} B^q_{is}B^s_{jk}-\\
&-L_jB^q_{ik}-\sum^n_{s\neq j} B^q_{js}B^s_{ik}-
\sum^n_{s\neq k}c^s_{ij} B^q_{sk}
\endaligned\tag2.22
$$
The derivatives $\L_i\varphi_i$, $\L_j\varphi_j$ and
$\L_k\varphi_k$ aren't defined by the equations \thetag{2.19}.
This gives the arbitrariness in $n$ functions for the solutions
of \thetag{2.19}. When they are nontrivial the relationships
\thetag{2.21} define the functional dependence between these
derivatives. Therefore they diminish the degree of arbitrariness.
In case of maximal arbitrariness the relationships \thetag{2.12}
should be trivial
$$
B^i_{jk}=B^j_{ik}=c^k_{ij} = 0
\text{, \ for } i\neq j, j\neq k, k\neq i \tag2.23
$$
{}From \thetag{2.23} due to the lemma 1 we get the
diagonalizability of the operator $\bold A(\bold u)$ by means of
point transformations from \thetag{1.3}. Due to \thetag{2.20}
the equations \thetag{2.19} are rewritten as follows
$$
L_i\varphi_j=-\frac{\lambda_{ji}\varphi_i-\lambda_{ij}\varphi_j}
{\lambda_i-\lambda_j}+c^j_{ij}\varphi_j
\text{, \ for } i\neq j\tag2.24
$$
The compatibility conditions for \thetag{2.24} are defined by
the quantities from \thetag{2.22} as $R^q_{kij}=0$. On taking
into account \thetag{2.23} these compatibility conditions are
exactly coincide with \thetag{2.13}. In spite of the fact that
the equations \thetag{2.10} and \thetag{2.24} are different
their compatibility conditions are the same and have the form of
semihamiltonity condition written in the frame of eigenvectors
of the operator $\bold A(\bold u)$. The above considerations
prove the following version of the theorem 1. \par
\proclaim{Theorem 5} System of the equations \thetag{1.1} with
the matrix of general position possess the continuous set of
conservation laws parameterized by $n$ functions of one variable
if and only if it is diagonalizable and semi-Hamiltonian.
\endproclaim
\head
3. The Froelicher-Nijenhuis bracket and the Lie superalgebra
   of vector-valued differential forms.
\endhead
     Let $\bold A$ be the tensor field of the type $(1,p)$ and let
it be skew symmetric in covariant components. Then $\bold A$
defines the vector-valued $p$-form $\bold A=\bold A(\bold X_1,
\dots, \bold X_p)$. Here $\bold  X_1,\dots, \bold X_p$ are some
arbitrary vector fields. Let $\bold B$ be the second tensor
field of the type $(1,q)$ which define vector valued $q$-form
$\bold B(\bold X_1,\dots,\bold X_q)$. We shall call $\bold A$ and
$\bold B$ the vector fields of rank 1 if they are of the following
form
$$
\bold A=\bold a\otimes\alpha
\hskip 5em plus 2em minus 2em
\bold B=\bold b\otimes\beta
\tag3.1
$$
where $\bold a$ and $\bold b$ are vector fields while $\alpha$
and $\beta$ are the differential forms. For the tensor fields of
the form \thetag{3.1} we define the pairing $\{\bold A,\bold B\}$
(it is known as Froelicher-Nijenhuis bracket)
$$
\aligned
\{\bold A,\bold B\}&=[\bold a,\bold b]\otimes\alpha\wedge
\beta - \bold a\otimes L_{\bold b}\alpha\wedge\beta +
\bold b\otimes \alpha\wedge L_{\bold a}\beta + \\
&+(-1)^p \bold a\otimes \iota_{\bold b}\alpha\wedge d\beta +
(-1)^p \bold b\otimes d\alpha\wedge\iota_{\bold a}\beta
\endaligned\tag3.2
$$
Via $\iota_{\bold a}$ and $\iota_{\bold b}$ in formula
\thetag{3.2} we denote the differentiations of substitution. For
the $r$-form $\omega$ and for the vector field $\bold c$ the
expression $\iota_{\bold c}\omega$ is a $r-1$-form
$$
\iota_{\bold c}\omega(\bold X_1,\dots,\bold X_{r-1})=
r\omega(\bold c,\bold X_1,\dots,\bold X_{r-1})
$$
The operation $\iota_{\bold c}$ is also known as inner product
with respect to the vector field $\bold c$ (see \cite{8}). \par
\proclaim{Theorem 6} The bracket $\{\bold A,\bold B\}$ defined
for the tensor fields $\bold A$ and $\bold B$ of rank 1 by the
formula \thetag{3.2} is uniquely continued for the arbitrary
tensor fields of the types $(1,p)$ and $(1,q)$ skew symmetric
in their covariant components. \endproclaim
     PROOF. Each tensor field of the type $(1,p)$ can be written
as a sum of tensor fields of rank one as follows
$$
\bold A=\sum_i\bold A_i=\sum_i\bold a_i\otimes
\alpha_i\tag3.3
$$
The analogous formula can be written for the field $\bold B$.
Therefore the bracket \thetag{3.2} for the arbitrary $\bold A$
and $\bold B$ can be redefined as
$$
\{\bold A,\bold B\}=\sum_i\sum_j\{\bold A_i,\bold B_j\}
\tag3.4
$$
However the expansion \thetag{3.3} is not unique. Therefore the
definition $\{\bold A,\bold B\}$ by means of \thetag{3.4} should
be tested for the correctness. The arbitrariness in the expansion
\thetag{3.3} for $\bold A$ is defined by the following identities
in tensor algebra
$$
\align
\split
&(\bold a + \tilde\bold a)\otimes\alpha =
\bold a\otimes\alpha + \tilde\bold a\otimes\alpha \\
&\bold a\otimes(\alpha + \tilde\alpha) =
\bold a\otimes\alpha + \bold a\otimes\tilde\alpha
\endsplit\tag3.5 \\
&(f \bold a)\otimes\alpha= \bold a\otimes(f\alpha)
\tag3.6
\endalign
$$
where $f$ is an arbitrary scalar field. The arbitrariness due to
\thetag{3.5} does not influence to the value of bracket
\thetag{3.4} since the relationship \thetag{3.2} is additive with
respect to $\bold a$ and $\alpha$. Let's ensure that the
arbitrariness due to \thetag{3.6} also doesn't make the influence
to the value of $\{\bold A, \bold B\}$. In order to do it we
calculate this bracket by \thetag{3.2} first for $\bold A= (f
\bold a)\otimes\alpha$ then for $\bold A=\bold a\otimes (f\alpha)$
and after all we compare the results. All these calculations are
based on the following formulae from \cite{8}
$$
\xalignat 2
[f\bold a,\bold b]&= f[\bold a,\bold b]-\bold a L_{\bold b}f
& L_{f\bold a}\beta&=f L_{\bold a}\beta + df\wedge
\iota_{\bold a}\beta\\
L_{\bold b}(f\alpha)&= fL_{\bold b}\alpha + L_{\bold b}f \alpha
& \iota_{f\bold a}\beta&=f\iota_{\bold a}\beta
\endxalignat
$$
Since these calculations are standard we did not write them here.
Theorem is proved $\square$\par
     Note that for $p=q=0$ the bracket \thetag{3.2} coincides
with the ordinary commutator of vector fields. For the arbitrary
values of $p$ and $q$ the algebraic properties of this bracket
are given by the following theorem. \par
\proclaim{Theorem 7} The bracket $\{\bold A,\bold B\}$ for the
tensor fields of rank 1 defined by \thetag{3.2} and then
generalized by \thetag{3.4} satisfies the relationships
$$
\align
&\{\bold A,\bold B\}+(-1)^{pq}\{\bold B,\bold A\}=0\\
&\{\{\bold A,\bold B\},\bold C\}(-1)^{rp} +
 \{\{\bold B,\bold C\},\bold A\}(-1)^{pq} +
 \{\{\bold C,\bold A\},\bold B\}(-1)^{qr}=0
\endalign
$$
because of which it defines the structure of graded Lie
superalgebra in tensor fields of type $(1,m)$ skew symmetric
in covariant components. \endproclaim
     Let $\bold A$ and $\bold B$ be tensor fields of the type
$(1.1)$ i.e. operator fields. Tensor field $\bold S=2\{\bold A,
\bold B\}$ is the vector valued 2-form. Its values may be
calculated by the following formula
$$
\aligned
&\bold S(\bold X,\bold Y)=[\bold A\bold X,\bold B\bold Y] +
[\bold B\bold X,\bold A\bold Y]+\\
&+\bold A\bold B[\bold X,\bold Y]+
 \bold B\bold A[\bold X,\bold Y] -
 \bold A[\bold X,\bold B\bold Y]-\\
&-\bold A[\bold B\bold X,\bold Y]-
\bold B[\bold X,\bold A\bold Y]-\bold B[\bold A\bold X,\bold Y]
\endaligned
\tag3.7
$$
Tensor $\bold S$ is known as the torsion of Nijenhuis for the
operator fields $\bold A$ and $\bold B$ (see \cite{5},
\cite{6} and \cite{8}). \par
\head
4. The construction of semihamiltonity tensor.
\endhead
     First let's recall the classical invariant criterion of
diagonalizability for the operator field $\bold A$. We mentioned
this criterion in section 1 (see also \cite{3}, \cite{4} and
\cite{6}). Let's consider the particular form of tensor
\thetag{3.7}
$$
\bold N=\{\bold A,\bold A\}
$$
It is usually called the tensor of Nijenhuis. From \thetag{3.7}
we obtain
$$
\aligned
\bold N(\bold X,\bold Y)&=[\bold A\bold X,\bold A\bold Y] +
\bold A^2[\bold X,\bold Y]-\\
&-\bold A[\bold X,\bold A\bold Y]-
\bold A[\bold A\bold X,\bold Y]
\endaligned
\tag4.1
$$
Tensor of Haantjes is defined via the tensor of Nijenhuis
\thetag{4.1} according to the formula
$$
\aligned
\bold H(\bold X,\bold Y)&=\bold N(\bold A\bold X,\bold A\bold Y)
+ \bold A^2\bold N(\bold X,\bold Y)-\\
&-\bold A\bold N(\bold X,\bold A\bold Y)-
\bold A\bold N(\bold A\bold X,\bold Y)
\endaligned
\tag4.2
$$
It has the same type as the tensor $\bold N$. It is also the
vector-valued skew symmetric 2-form. \par
\proclaim{Theorem 8 {\rm (criterion of diagonalizability)}}
The operator $\bold A(\bold u)$ of general position with mutually
distinct eigenvalues is diagonalizable by means of point
transformations \thetag{1.3} if and only if its tensor of
Haantjes \thetag{4.2} is identically zero. \endproclaim
     The criterion of diagonalizability in form of this theorem
was first proved in \cite{3}. It was applied to the systems of
equations \thetag{1.1} in \cite{4}. \par
     PROOF. Because of skew symmetry of bilinear form
\thetag{4.2} it's enough to test vanishing this form only for
vector fields $\bold X = \bold X_i$ and $\bold Y = \bold X_j$
from the frame of eigenvectors of the operator $\bold A(\bold u)$
with $i\neq j$. By direct calculations we obtain
$$
\bold H(\bold X_i,\bold X_j)=
\sum^n_{k=1}(\lambda_i-\lambda_k)^2
(\lambda_j-\lambda_k)^2 c^k_{ij} \bold X_k
\tag4.3
$$
Because of \thetag{4.3} the equality $\bold H(\bold X_i,
\bold X_j)=0$ is equivalent to \thetag{2.12}. Then we are only
to apply the lemma 1. Criterion is proved. $\square$\par
     Practical use of this criterion for the testing the
diagonalizability is based on the following formulae for
the components of tensors $\bold N$ and $\bold H$ expressing
them through the components of matrix $\bold A(\bold u)$
$$
\align
&N^k_{ij}=\sum^n_{s=1}\bigl(A^s_i\partial_sA^k_j-
A^s_j\partial_sA^k_i+A^k_s\partial_jA^s_i-
A^k_s\partial_iA^s_j\bigr)\hskip -5em\tag4.4\\
\split
&H^k_{ij}=\sum^n_{s=1}\sum^n_{r=1}\bigl(
A^k_sA^s_rN^r_{ij}-\\
&\hskip 4em -A^k_sN^s_{rj}A^r_i-A^k_sN^s_{ir}A^r_j+
N^k_{sr}A^s_iA^r_j
\bigr)
\endsplit\tag4.5
\endalign
$$\par
     Let $\bold B$ be the operator field i.e the tensor field of
type $(1,1)$ and let $\bold Q$ be the skew symmetric tensor field
of type $(1,2)$. Through $\bold K$ we denote the
Froelicher--Nijenhuis bracket of these two fields $\bold K=
3\{\bold Q,\bold B\}$. For the 3-form $\bold K$ we have
$$
\aligned
&\bold K(\bold X,\bold Y,\bold Z) =
  \bold B[\bold X,\bold Q(\bold Y,\bold Z)] -
  [\bold B\bold X,\bold Q(\bold Y,\bold Z)] + \\
&+\bold B\bold Q(\bold X,[\bold Y,\bold Z]) +
  \bold Q(\bold X,\bold B[\bold Y,\bold Z]) -
  \bold Q(\bold X,[\bold B\bold Y,\bold Z]) - \\
&-\bold Q(\bold X,[\bold Y,\bold B\bold Z]) +
  \dots
\endaligned\tag4.6
$$
Dots in \thetag{4.6} denote 12 summand that can be obtained
from explicitly written summands in \thetag{4.6} by means of
the cyclic transposition of $\bold X$, $\bold Y$ and $\bold Z$.
\par
     Starting with deriving the invariant semihamiltonity
criterion we should note that everywhere above (see theorems 1,
3 and 5) the semihamiltonity comes together with
diagonalizability. It doesn't play the separate role. Therefore
we will use the following scheme of action: we will construct the
tensor vanishing of which gives the equations \thetag{1.4} after
bringing this tensor to the coordinates where the matrix
$\bold A(\bold u)$ is diagonal. The equations \thetag{1.4} are
rational. Let's rewrite them in polynomial form. In order to do
it we introduce the following quantities
$$
\aligned
\alpha^k_{kij}&= - (\lambda_i-\lambda_j)(\lambda_i-\lambda_k)
(\lambda_j-\lambda_k)\partial_{ij}\lambda_k - \\
&-(\lambda_i-\lambda_j)(\lambda_i+\lambda_j-2\lambda_k)
\partial_i\lambda_k\partial_j\lambda_k + \\
&+(\lambda_i-\lambda_k)^2\partial_i\lambda_j\partial_j\lambda_k -
(\lambda_j-\lambda_k)^2\partial_j\lambda_i\partial_i\lambda_k
\endaligned
\tag4.7
$$
The semihamiltonity \thetag{1.4} then is written as the condition
of vanishing the quantities \thetag{4.7}
$$
\alpha^k_{kij}=0\text{, \ for } i\neq k\neq j\tag4.8
$$
Partial derivatives in \thetag{4.7} as in \thetag{1.4} are
calculated with respect to the variables $u^1,\dots,u^n$ for
which the matrix $\bold A(\bold u)$ is diagonal. The frame of
eigenvectors $\bold X^1,\dots,\bold X^n$ is chosen to be the
coordinate frame for such variables. \par
     Using the Froelicher--Nijenhuis bracket we construct the
tensor $\bold K$ from the matrix $\bold A$ as follows
$$
\bold K = 3\{\{\bold A,\bold A\},\bold A^2\}=
3\{\bold N,\bold A^2\} \tag4.9
$$
Tensor $\bold K$ defines the vector-valued 3-form the values
of which for the frame vectors can be calculated according to
\thetag{4.6}. As a result from \thetag{4.9} we have
$$
\bold K(\bold X_k,\bold X_i,\bold X_j)=
K^k_{kij}\bold X_k +
K^i_{kij}\bold X_i +
K^j_{kij}\bold X_j
\tag4.10
$$
(there is summation by $i,j,k$ here). We are interested only in
one group of components of tensor $\bold K$. Others can be
obtained by cyclic transpositions of indices in $K^k_{kij}$. The
values of components from this group in \thetag{4.10} are defined
by the following formula
$$
\aligned
K^k_{kij}-\alpha^k_{kij}=&2(\lambda_i-\lambda_j)
\bigl[(\lambda_i-\lambda_k)+(\lambda_j-\lambda_k)\bigr]\\
&(\partial_i\lambda_k\partial_j\lambda_k-
\partial_i\lambda_j\partial_j\lambda_k-\partial_j\lambda_i\partial_i
\lambda_k)\endaligned\tag4.11
$$
{}From \thetag{4.11} we see that the difference $K^k_{kij}-
\alpha^k_{kij}$ contains only the derivatives of the first order.
Now taking the tensor $\bold N$ we construct another tensor
$\bold M$. Corresponding polylinear form is the following
$$
\aligned
&\bold M(\bold X,\bold Y,\bold Z)=
        \bold N(\bold X,\bold A\bold N(\bold Y, \bold Z))+\\
&+\bold N(\bold A \bold X,\bold N(\bold Y,\bold Z))
        - \bold N(\bold N(\bold X,\bold Z),\bold A \bold Y)+\\
&+\bold N(\bold N(\bold X,\bold Y),\bold A \bold Z)
        - \bold N(\bold X, \bold N(\bold A \bold Y,\bold Z)) -\\
&-\bold N(\bold X,\bold N(\bold Y,\bold A \bold Z))
\endaligned
\tag4.12
$$
For the tensor $\bold M$ from \thetag{4.12} computed in the frame
of eigenvectors of the operator $\bold A(\bold u)$ we have
$$
\bold M(\bold X_k,\bold X_i,\bold X_j)=
M^k_{kij}\bold X_k +
M^i_{kij}\bold X_i +
M^j_{kij}\bold X_j
\tag4.13
$$
For the of components components of $\bold M$ in \thetag{4.13}
we get
$$
\aligned
M^k_{kij}=-&(\lambda_i-\lambda_j)(\lambda_i-\lambda_k)(\lambda_j-\lambda_k)\\
&(\partial_i\lambda_k\partial_j\lambda_k-
\partial_i\lambda_j\partial_j\lambda_k-\partial_j\lambda_i\partial_i
\lambda_k)\endaligned\tag4.14
$$
The coefficients $M^i_{kij}$ and $M^j_{kij}$ aren't of interest
for us now. Comparing \thetag{4.11} with \thetag{4.14} we define
another tensor $\bold Q$
$$
\aligned
&\bold Q(\bold X,\bold Y,\bold Z)=
         \bold K(\bold A\bold X,\bold A\bold Y,\bold Z)
        - \bold K(\bold A^2\bold X,\bold Y,\bold Z)-\\
&- \bold K(\bold X,\bold A\bold Y,\bold A\bold Z)
        + \bold K(\bold A\bold X,\bold Y,\bold A\bold Z)
        + 4 \bold M(\bold A\bold X,\bold Y,\bold Z)-\\
&- 2 \bold M(\bold X,\bold A\bold Y,\bold Z)
        - 2 \bold M(\bold X,\bold Y,\bold A\bold Z)
\endaligned\tag4.15
$$
For the components of this tensor in the frame of eigenvectors of
operator $\bold A(\bold u)$ we get the relationship
$$
\bold Q(\bold X_k,\bold X_i,\bold X_j)=
Q^k_{kij}\bold X_k +
Q^i_{kij}\bold X_i +
Q^j_{kij}\bold X_j
\tag4.16
$$
which is analogous to \thetag{4.10} and \thetag{4.13}. The
components in \thetag{4.16} which are of interest for us are
expressed through \thetag{4.7}. They have the form
$$
Q^k_{kij}=-(\lambda_i-\lambda_k)(\lambda_j-\lambda_k)
\alpha^k_{kij}\tag4.17
$$
Now on a base of \thetag{4.17} we are able to construct the
semihamiltonity tensor which is the main goal of the whole
paper. It is defined by the following formula
$$
\aligned
&\bold P(\bold X,\bold Y,\bold Z)=
        \bold A\bold Q(\bold X,\bold A \bold Y,\bold Z)+\\
&+ \bold A\bold Q(\bold X,\bold Y,\bold A \bold Z)
        - \bold A^2\bold Q(\bold X,\bold Y,\bold Z)
        - \bold Q(\bold X,\bold A \bold Y,\bold A \bold Z)
\endaligned
\tag4.18
$$
It is easy to check that for arbitrary three vectors from the
frame $X^1,\dots,X^n$ one has the relationship
$$
\bold P(\bold X_k,\bold X_i,\bold X_j)=
(\lambda_i-\lambda_k)^2 (\lambda_j-\lambda_k)^2
\alpha^k_{kij}\bold X_k\tag4.19
$$
\par
As a result we proved the following theorem (the invariant
criterion of semihamiltonity).\par
\proclaim{Theorem 9} The diagonalizable operator of general
position $\bold A$ with mutually distinct eigenvalues is
semi-Hamiltonian if and only if when associated tensor $\bold P$
from \thetag{4.18} is identically zero. \endproclaim
     The proof of this theorem follows directly from \thetag{4.8}
and \thetag{4.9}. It doesn't require any comments. Concluding all
above considerations we give the formulae which enable us to
calculate the components of the tensor of semihamiltonity
$\bold P$ from the matrix $\bold A(\bold u)$
$$
\aligned
P^s_{kij}=\sum^n_{p=1}\sum^n_{q=1}&\bigl(
   A^s_p Q^p_{kqj} A^q_i + A^s_p Q^p_{kiq} A^q_j - \\
&- A^s_q A^q_p Q^p_{kij} - Q^s_{kpq} A^p_i A^q_j\bigr)
\endaligned\tag4.20
$$
This formula is derived from \thetag{4.18}. Components of the tensor
$\bold Q$ in the formula \thetag{4.20} are calculated on a base of
\thetag{4.15}
$$
\align
&Q^s_{kij} = \sum^n_{p=1}\sum^n_{q=1} \bigl(
          A^p_k K^s_{pqj} A^q_i +
          A^p_k K^s_{piq} A^q_j -
          A^p_q A^q_k K^s_{pij} - \\
       &- K^s_{kpq} A^p_i A^q_j \bigr) +
\sum^n_{p=1} \bigl(
        4 A^p_k M^s_{pij} -
        2 M^s_{kpj} A^p_i -
        2 M^s_{kip} A^p_j \bigr)
\endalign
$$
Components of $\bold M$ in the above formula are found from
\thetag{4.12}
$$
\align
&M^s_{kij}=\sum^n_{p=1}\sum^n_{q=1} \bigl(
    N^s_{kp} A^p_q N^q_{ij}
  + N^s_{pq} A^p_k N^q_{ij} - \\
& - N^s_{pq} N^p_{ik} A^q_j
  - N^s_{pq} N^p_{kj} A^q_i
  - N^s_{kp} N^p_{iq} A^q_j
  - N^s_{kp} N^p_{qj} A^q_i \bigr)
\endalign
$$
Tensor $\bold K$ are computed through tensor of Nijenhuis with
the use of the bracket of Froelicher and Nijenhuis on a base of
formula \thetag{4.9}. Let's take $\bold B=\bold A^2$. Then for
the components of $\bold K$ we have
$$
\aligned
K^s_{kij} &= \sum^n_{p=1}\bigl(
  B^s_p\partial_k N^p_{ij}-B^p_k\partial_p N^s_{ij}+\\
&+N^p_{ij}\partial_p B^s_k - N^s_{kp}\partial_i B^p_j
 +N^s_{kp}\partial_j B^p_i \bigr) + \dots
\endaligned\tag4.21
$$
By dots in \thetag{4.21} we denote 10 summand that can be
obtained from 5 explicit summands by cyclic transposition of
indices $i$, $j$ and $k$. \par
     The above formulae huge enough for direct calculations.
However modern computer systems for analytical calculations
solve this problem for any particular equations of hydrodynamical
type in applications.

\enddocument